\documentclass[useAMS,usenatbib,usedcolumn]{mn2e}

\input ifpdf.sty

\ifpdf
  \usepackage[pdftex,colorlinks,urlcolor=blue,draft]{hyperref}
\else
  \usepackage{url}
\fi

\usepackage{graphicx}
\usepackage{relsize}
\usepackage{amsbsy} 
\usepackage{amssymb}
\usepackage{amsmath}

\newcommand{\unit}[1]{\,\mathrm{#1}}

\newcommand{\MVrbin}{M_{{\rm V},r_{\rm bin}}}
\newcommand{\MVrhalf}{M_{{\rm V},r_{1/2}}}

\newlength{\wminus}
\newcommand{\subox}{{\scshape Superbox}}

\begin{document}
\title[%
Are dSph galaxies of tidal origin? (acc. for pub.)
]{Dwarf-spheroidal satellites: are they of tidal origin?}

\author[M. Metz, P. Kroupa]
{Manuel Metz, Pavel Kroupa\thanks{E-mail: \protect\url{pavel@astro.uni-bonn.de}}\\
Argelander-Institut f\"ur Astronomie\thanks{Founded by merging of the
\emph{Sternwarte}, \emph{Radioastronomisches Institut}, and \emph{Institut f\"ur Astrophysik
und Extraterrestrische Forschung der Universit\"at Bonn}}, Universit\"at
Bonn, Auf dem H\"ugel 71, D--53121 Bonn, Germany
}

\date{Accepted 2006 December 21 / Received 2006 November 14}
\pagerange{\pageref{firstpage}--\pageref{lastpage}} \pubyear{2006}

\maketitle
\label{firstpage}

\begin{abstract}
The Milky Way and Andromeda must have formed through an initial epoch of sub-structure merging. As a result of fundamental physical conservation laws tidal-dwarf galaxies (TDGs) have likely been produced. Here we show that such TDGs appear, after a Hubble-time of dynamical evolution in the host dark-matter halo, as objects that resemble known dSph satellite galaxies. We discuss the possibility that some of the Milky Way's satellites may be of tidal origin.
\end{abstract}


\section{Introduction}\label{sec_intro}
Both large spirals of the Local Group, the Milky Way (MW) and Andromeda (M31), have faint satellite galaxies within their virial radii. The number of known companions steadily increased in the last two years, since more, very faint dwarf galaxies were detected with the help of the Sloan Digital Sky Survey \citep[SDSS,][]{york00}: six newly discovered companions of the Milky Way \citep{willm05,belok06a,zucke06} and five of Andromeda \citep{zucke04,zucke06b,marti06} were classified as dwarf spheroidal galaxies (dSph). The dSphs are by far the most numerous galaxies in the Local Group and the Universe as a whole, populating the very faint end of the galaxy luminosity function. Presently known, the MW has 19 and M31 has 12 dSph satellites. Their spatial distribution is disc-like (certainly for the 11 brightest MW satellites; probably for M31) posing a challenge to understanding their origin (\citealp*{kroup05}; \citealp{zentn05,libes05,yang06,koch06}; \citealp*{metz06}) and suggesting correlated orbital angular momenta (\citealp{lynde82}; \citealp*{palma02}).

Traditionally, dSph galaxies have mostly been thought to be dark-matter dominated objects, primordial subhaloes that captured enough gas to form stars and that they appear today as very faint, extended and dark-matter dominated objects (e.g., \citealp{white78,stoeh02}; \citealp*{read06}). As such they would be cosmological hierarchical building blocks, subhaloes that have not yet merged with their host haloes.
The theoretical treatments differ though rather substantially, and correct predictions have never emerged so far. The general approach is to perform hierarchical structure-formation calculations combined with an algorithm describing the dark-matter -- luminous-matter bias, tuned to explain currently available data \citep[e.g.,][]{libes05,moore06}. Inferred dark-matter profiles are, however, not consistent with the cold-dark-matter (CDM) concordance cosmological hypothesis \citep{wilki06,gilmo06}. A possible alternative would be a warm dark-matter cosmology \citep[e.g.,][]{moore00} or an alternative gravitational theory \citep[e.g.,][]{brada00}. Based on their chemical enrichment, today's dSph galaxies are also excluded as major contributors to the Galactic stellar halo, since they have been born from pre-enriched material (\citealp*{unava96}; \citealp{helmi06}).

A different possible origin for dwarf galaxies arises from fundamental physical processes, conservation of energy and angular momentum: the formation of tidal-dwarf galaxies (TDGs), already pointed out by \citeauthor{zwick56} in 1956, a process that is observed to occur in the local (e.g.,~\citealp*{hunsb96}; \citealp{duc98}; \citealp*{weilb03,walte06}) and distant Universe \citep[e.g.,][]{strau06}. Interacting gas-rich galaxies throw out long, thin tidal arms of gaseous and stellar material which subsequently fragments, collapses and starts forming new stars. High-resolution images show young TDG candidates to consist of star-cluster complexes \citep{jarre06}, a star-formation mode typical for intense star-bursts as is for example on-going in the Antennae galaxies \citep{whitm05}. Such cluster complexes typically survive the first few hundred~Myr \citep{kroup98,fellh02a,fellh05}, and can evolve into dSph-like objects currently known \citep{kroup97}. 

The formation of TDGs is evident in simulations of gas-rich galaxy encounters as well (\citealp{barne92}; \citealp*{elmeg93}; \citealp{bourn06}; \citealp*{wetzs05}). Recently \citet{bourn06} studied the formation and survival of TDGs produced in encounters of colliding galaxies with different mass ratia and encounter parameters of the progenitors using a particle-mesh code. The gaseous component was represented by sticky particles. They found that about one quarter of their TDG candidates were `long lived' objects ($>$~2~Gyrs), which were predominately formed at the outermost regions of the tidal tails, and claimed that up to 10 per~cent of the dwarf galaxy population may be of tidal origin.
In their high-resolution simulations, including gas dynamics using a smoothed particle hydrodynamics (SPH) code, \citet{wetzs05} showed that TDGs are dominated by the gas; dissipational hydrodynamics is crucial for their formation. Some TDGs may become intergalactic spam \citep{elmeg93} appearing as ordinary dIrr galaxies \citep*{hunte00}. 
TDGs may significantly contribute to dwarf galaxy populations in dense environments: 
\citet{hunsb96} suggested that about half of the current dwarf galaxy population in compact groups could be of tidal origin, and \citet{okaza00} demonstrated that all dE galaxies in clusters can be of tidal origin under standard cosmological structure formation conditions.

All simulations of tidal-dwarf formation predict low true mass-to-light $(M/L)_{\rm true}$ ratia, which is also evident from observations \citep{walte06}. A low $(M/L)_{\rm true}$ ratio seems to contradict the observed mass-to-light ratia of the dSph satellites of the Milky Way, excluding them to be of tidal origin. However, it has already been shown that satellite galaxies may be out of virial equilibrium: due to close peri-galactic passages they can be significantly perturbed but not fully disrupted \citep{kroup97,kless98,fleck03}. Such objects show a number of properties observed for Local Group dSph satellites, like internal substructures and irregular shapes. Recent observations also support the picture that tidal interaction may play a key role in the evolution of satellite galaxies. \citet{sohn06} discoverd extra tidal features in Leo~I, located at a distance of $254\unit{kpc}$ from the MW centre.

In this paper we present new results extending the results of \citet{kroup97}. In Section~\ref{sec_data} the simulations are described and the results are presented in Section~\ref{sec_results}. In Section~\ref{sec_discuss} we discuss the results and give some concluding remarks in Section~\ref{sec_conclusions}.

\section{Satellite galaxy models}
\subsection{The data}\label{sec_data}
The simulations were carried out in an analog manner as in \citet{kroup97}: satellite galaxies were set-up as Plummer-spheres, having masses of $10^7 \unit{M_{\sun}}$ \emph{without dark-matter} and initial absolute magnitudes $M_{\rm V}=-11.5 \unit{mag}$ being represented by $3 \times 10^5$ particles. These values correspond to a true mass-to-light ratio $(M/L)_{\rm true}=3$ for each particle, and the models represent TDGs that have blown out their gas within the first Gyr after their formation \citep{recch06}. They were allowed to relax before being injected into the host halo with an isothermal profile of total mass of $2.85 \times 10^{12} \unit{M_{\sun}}$, core-radius $5\unit{kpc}$ and cutoff radius $250\unit{kpc}$ with circular velocity $v_{\rm c}=220 \unit{km\,s^{-1}}$. The satellite galaxies started at a distance of $60\unit{kpc}$ and $100 \unit{kpc}$ from the centre of the host halo and were given different initial velocities leading to different eccentricities of their orbits (see Table~\ref{tab_props}). The evolution was followed in a live halo. Every satellite evolves by periodically losing matter at its peri-galactic passages \citep{piate95} until a stage was reached when the original satellite was nearly disrupted and has evolved into a quasi-stable remnant. For a detailed description of the initial set-ups of the satellite galaxies and the host halo see \citet{kroup97}. The integration was done using the particle-mesh code \subox\ \citep{fellh00}. Properties of the satellite galaxies were compared with a direct N-body code \citep{kless98}, showing that the results agree well.

\begin{table}
\caption{Properties of simulations: in the first column an identifier is given. The second column lists the total integration time, the third the initial galactocentric distance. In the forth and fifth columns the mean peri-galactic distance and the mean eccentricity of the orbits are given, respectively.}\label{tab_props}
{
  \centering
  \begin{tabular}{lcccc}
  simulation & $t$ & $r_{init}$ & $\overline{r}_{\rm peri}$ & $\overline{ecc}$ \\
  & [Gyr] & [kpc] & [kpc] & \\
  \hline
  RS1-4     &  8.8 & 100 & 15.9 & 0.72 \\
  RS1-5     & 10.7 & 100 & 24.6 & 0.59 \\
  RS1-32    &  5.5 & 100 &  3.9 & 0.92 \\
  RS1-64    &  5.0 & 100 &  4.1 & 0.92 \\
  RS1-109   &  8.8 &  60 &  9.1 & 0.74 \\
  RS1-113   &  6.6 &  60 & 22.4 & 0.46 \\
  RS1-24-32 & 11.0 &  60 & 25.6 & 0.40 \\
  RS1-24-64 &  8.2 &  60 & 22.8 & 0.44 \\
  \end{tabular}
}
\end{table}

Observational parameters for the modelled satellite galaxies are derived in an automated manner as a virtual observer located in the disc would do: member stars are photometrically selected based on their projected position on the virtual sky. The line-of-sight velocity dispersion is derived within the half-light radius from which an apparent mass-to-light ratio $(M/L)_{\rm obs}$ is calculated under the assumption that the satellite galaxy is in virial equilibrium. The statistical routines from \citet*{beers90} were employed to ensure the velocity outliers are removed from these calculations. The absolute magnitudes $\MVrbin$ and $\MVrhalf$ are determined within a fixed projected distance $r_{\rm bin}=1.5\unit{kpc}$ and within the projected half-light radius, respectively. For a complete description and a detailed analysis of individual parameters see \citet{kroup97}.

\subsection{Results}\label{sec_results}
\begin{figure}
  \resizebox{\hsize}{!}{
    \includegraphics{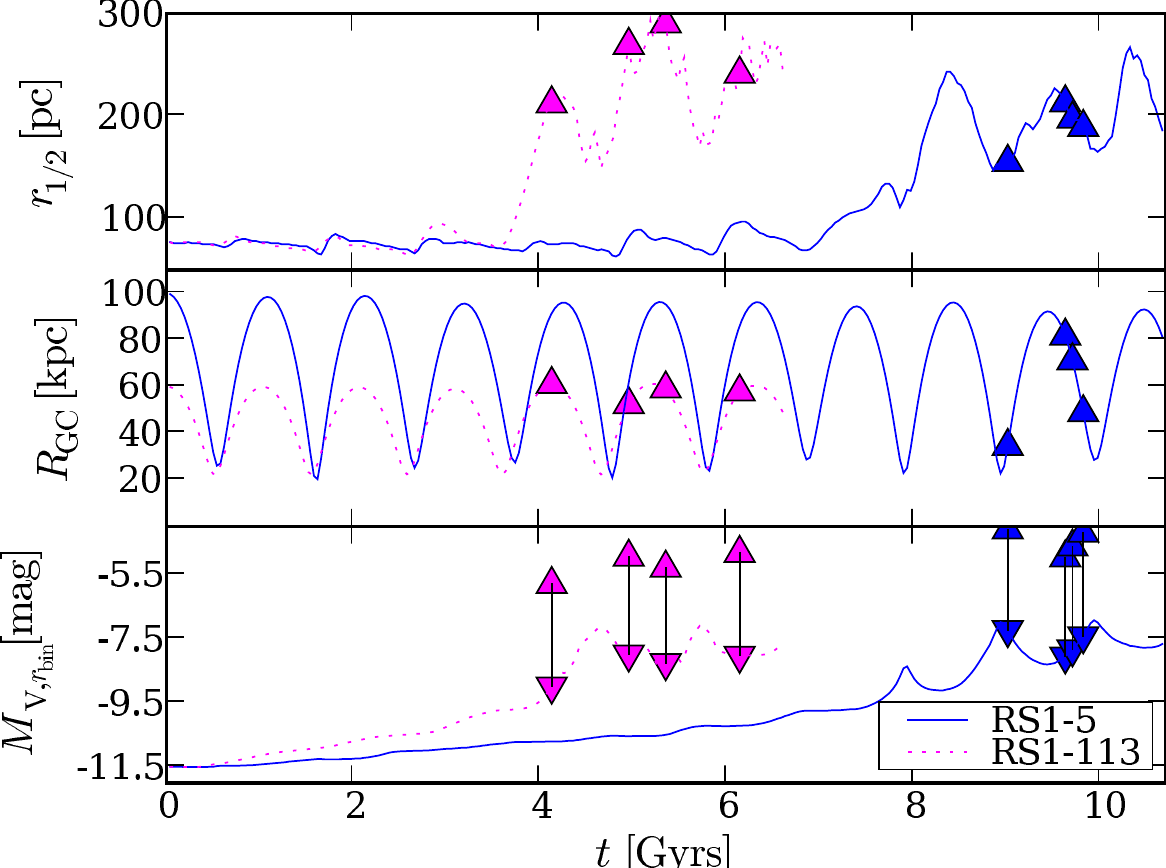}
  }
  \caption{The time evolutions of half-light radius, galactocentric distance and absolute magnitude $\MVrbin$ of two simulated satellite galaxies, models RS1-5 (solid curves) and RS1-113 (dotted curves). The triangle symbols mark snapshot values derived at random time-steps during the last two completely simulated orbits. For the snapshots of $\MVrbin$, the absolute magnitudes within the half-light radius are given in addition, marked with triangles pointing upwards, the corresponding values are connected by the thin solid lines.}
  \label{fig_triple}
\end{figure}

During their evolution the satellite galaxies change their apparent properties as determined by an observer. In Figure~\ref{fig_triple} the time evolution of the half-light radii and the apparent absolute magnitudes of two simulated satellites, RS1-5 and RS1-113, are shown. In the middle panel we additionally show the galactocentric distance of the satellites. Four snapshots, marked by the triangle symbols in Figure~\ref{fig_triple}, were taken for each simulation at random time-steps during the last two complete orbits of each simulation. This ensures that we are not biased in the selection of the snapshots and the probability to `observe' a satellite at a particular evolutionary phase (after the disruption) is proportional to the life-time of this phase.
As \citet{kroup97} already noted, after the disruption of the satellite galaxies a significant fraction of stars contributing to $\MVrbin$ of the quasi-stable remnant are `extra-tidal' stars. As such $\MVrbin$ is most likely over-estimated. Therefore we also plot the absolute magnitudes $\MVrhalf$ within the half-light radii, marked by triangles pointing upwards, which may be considered as the lower limit for the absolute luminosities.

\begin{figure*}
  \resizebox{16cm}{!}{
    \includegraphics{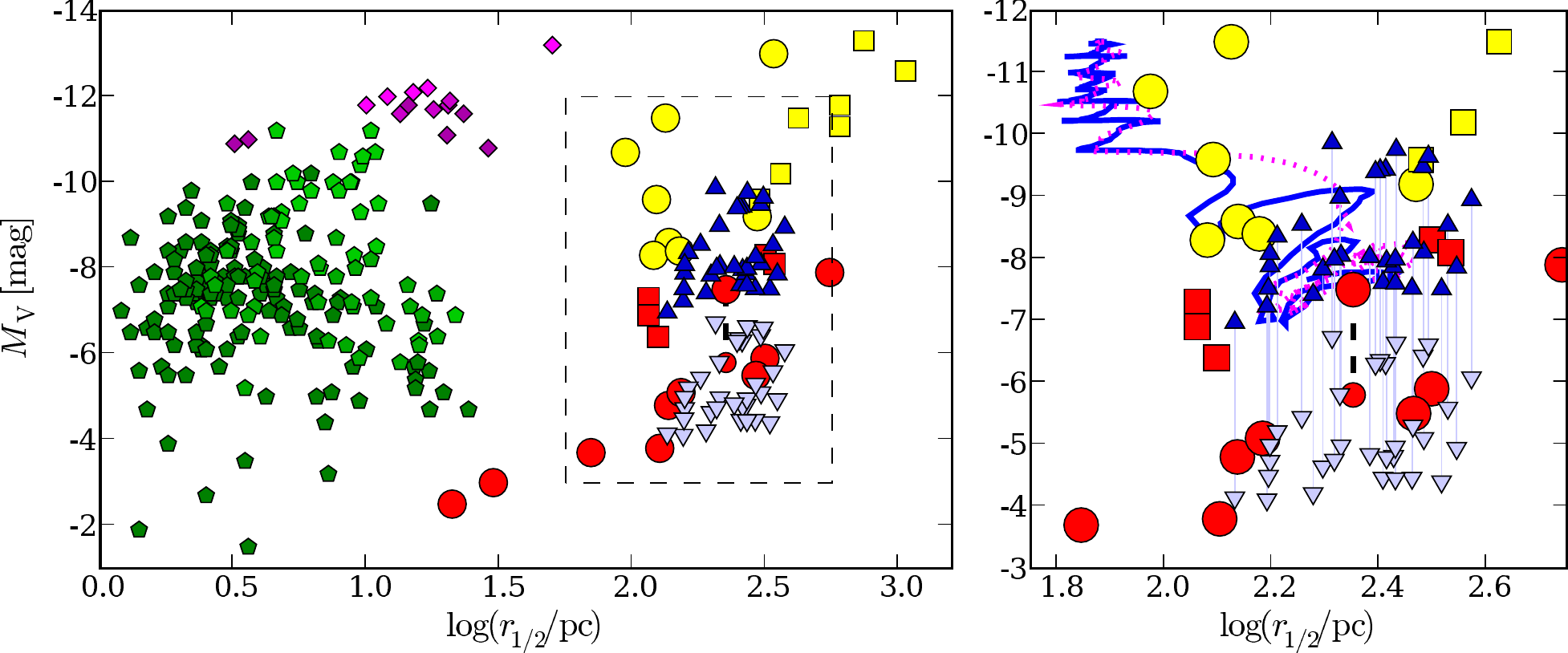}
  }
  \caption{Absolute magnitude versus half-light radius for Globular Clusters (pentagrams), ultra compact dwarf galaxies (diamonds), Milky Way (circles) and Andromeda (squares) dSph satellite galaxies, and simulated satellite galaxies (triangles). For the observational data of the MW and M31, the dark symbols mark the recent additions to the known list of companions from the SDSS, while light symbol mark the longer-known dSph's. For the MW dSph galaxy in Bo\"otes two values for the absolute magnitude are given (see text), the values are connected by a thick, dashed line. For the simulated data absolute magnitudes derived within a fixed projected distance $r_{\rm bin}$ (dark triangles) and within the variable distance $r_{1/2}$ (light triangles) are shown. In the right panel, the region in the left panel marked by the dashed rectangle is enlarged. Corresponding absolute magnitude values for the simulated satellites are connected by light solid lines. The evolutionary tracks of the models RS1-5 (solid curve) and RS1-113 (dotted curve) are shown.}
  \label{fig_hlr}
\end{figure*}

In Figure~\ref{fig_hlr} the absolute magnitude versus the half-light radius is shown for observational data: MW (circles) and M31 (squares) dSph satellite galaxies. The same data are used as in \citet{belok06a}, their figure 8 (Zucker, priv.\ comm.). For the four snapshots of each simulated satellite we show the two extreme values of the absolute magnitudes, $M_{{\rm V},r_{\rm bin}}$ (triangles pointing upwards) and $M_{{\rm V},r_{1/2}}$ (triangles pointing downwards), respectively. The corresponding values are connected by a thin solid line in the right panel. For the dSph in Bo\"otes \citep[Boo:][]{belok06} two values for the absolute magnitude are given, connected by a thick, dotted line. The second, more luminous value is taken from \citet{munoz06a} which was derived based on a comparison with UMa. In addition the development tracks of two simulated satellites (RS1-5 and RS1-113) are shown in the right panel. The tracks start at the upper-left and evolve to the lower-right direction as the satellites become more extended and less luminous.
It follows that all models evolve to the parameter region occupied by the observed satellites.

\begin{table}
  \caption{Observational properties of Local Group dSph galaxies. Where available in the literature we also give error estimates. References for $M_V$ and $(M/L)_{\sun}$ are given in the last column, if two references are given, the first is for the absolute magnitude and the second for the mass-to-light ratio (see Figure~\ref{fig_m2l}).}
  \label{tab_obsdata}
  {\centering
    \begin{tabular}{lccc@{}l}
    Name & $D$ [kpc] & $M_V$ [mag] & $(M/L)_{\sun}$ \\
    \hline
    UMi    &  50.2 &  -8.9 & 250        & $^{a,g}$\\
    Boo    &  57.6 &  -7.5 & 130        & \\
           &       &  -5.8 & 610        & $^{b}$ \\
    Scl    &  79.2 & -11.1 & 3          & $^{a}$ \\
    Dra    &  82.0 &  -8.8 & 440        & $^{a,h}$ \\
    Sex    &  89.2 &  -9.5 &  39        & $^{a}$ \\
    Car    & 102.7 &  -9.3 & $41^{+40}_{-25}$ & $^{c}$ \\
    UMa    & 104.9 & -6.75 & 550        & $^{d}$ \\
    For    & 140.1 & -13.2 & 4.4        & $^{a}$ \\
    Leo~II & 207.7 &  -9.6 & 17         & $^{a}$ \\
    Leo~I  & 254.0 & -12.3 & $5.3^{+1.6}_{-1.6}$ & $^{e}$\\
    And~IX &  42   &  -8.3 & $93^{+120}_{-50}$ & $^{f}$\\
    And~II & 185   & -11.3 &  $21^{+14}_{-10}$ & $^{g}$\\
    \end{tabular}
  }
  {\newline\footnotesize References: $^{a}$~\citet{mateo98}; $^{b}$~\citet{munoz06a}; $^{c}$~\citet{munoz06}; $^{d}$~\citet{kleyn05}; $^{e}$~\citet{sohn06}; $^{f}$~\citet{chapm05}; $^{g}$~\citet{cote99}}; $^{h}$~\citet{evans05}
\end{table}

\begin{figure*}
  \resizebox{10.5cm}{!}{
    \includegraphics{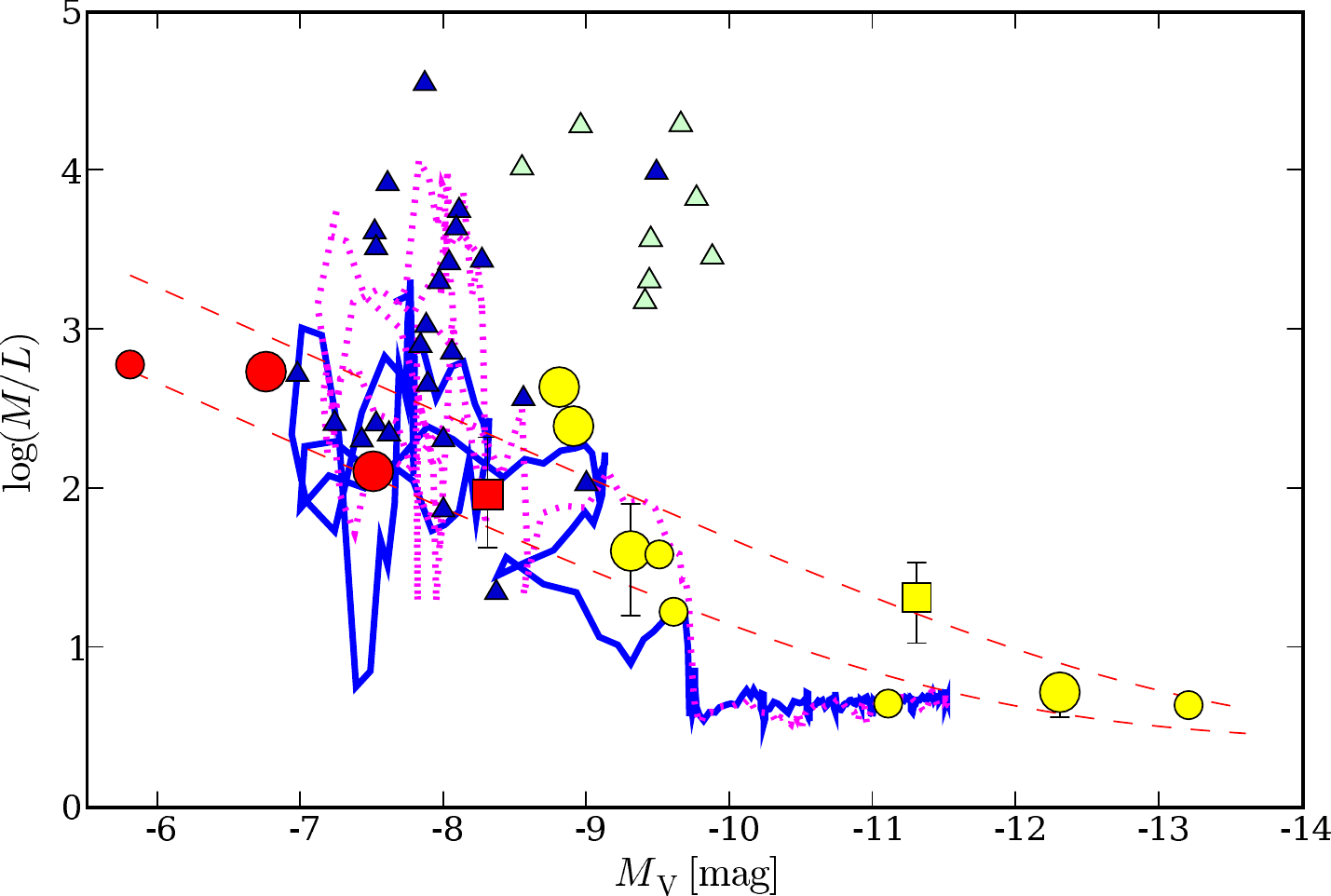}
  }
  \caption{Mass-to-light ratio versus absolute magnitude for MW and M31 dSph satellite galaxies, for references see Table~\ref{tab_obsdata}; symbols are as in Figure~\ref{fig_hlr}. For the MW, the data marked with smaller light circles were taken from \citet{mateo98}, others from more recent studies. The TDG models are shown as triangles using the same symbols as in the previous figures. See text for further details.}
  \label{fig_m2l}
\end{figure*}

Figure~\ref{fig_m2l} shows mass-to-light ratio (in solar units) versus absolute magnitude $M_{\rm V}$ of the observed Local Group dSph satellite galaxies and for the snapshots of the simulated satellites ($\MVrbin$), symbols being as in Figure~\ref{fig_hlr}. For the MW smaller, light symbols mark data taken from \citet{mateo98} that were determined from central velocity dispersions. The larger symbols mark more recent values as given in Table~\ref{tab_obsdata}; for Boo symbols are taken as in Figure~\ref{fig_hlr}. The dashed lines show the expected mass-to-light ratio if stars were embedded in dark-matter haloes of masses $1\times 10^7 \unit{M_{\sun}}$ and $4 \times 10^7 \unit{M_{\sun}}$, respectively \citep[see also][]{mateo98}.

The evolution tracks shown here for the simulated satellites now start at the lower-right end. In the course of time the satellites get fainter and more disturbed, leading to an increase in the apparent $(M/L)_{\rm obs}$ while the particle value is constant $(M/L)_{\rm true}=3\; (M/L)_{\sun}$.

\section{Discussion}\label{sec_discuss}
In Figures~\ref{fig_triple} -- \ref{fig_m2l} it can be seen that dwarf satellite galaxies without dark-matter orbiting in massive host haloes span a wide range of observed properties. Energy is pumped into the satellite during peri-galactic passages because of the interaction with the Galactic tidal field leading to an expansion of the satellite. After the disruption of the initial galaxy, a quasi-stable remnant remains, that matches many of the observed properties of today's dSph satellite galaxies of the Milky Way and Andromeda \citep{kroup97}. While initially being identical objects, after several orbits the remnants can appear as objects having very different observed properties, depending on the eccentricities and peri-galactic distance of their orbits. Also during their life-time individual satellites may appear as having different properties during disruptive and subsequent relaxation phases.

Figure~\ref{fig_hlr} shows the simulated satellite galaxies to have similar half-light radii and absolute magnitudes as the Local Group dSph satellites, but to be about one or two magnitudes too bright compared to the latest additions from the SDSS. As the luminosity within $r_{\rm bin}$ of the simulated satellites is likely to be overestimated, the true absolute luminosity a real observer would determine is likely to be somewhat lower. On the other hand, it is not clear whether the observed absolute luminosities of some of the very recently discovered dSph companions of the MW are not underestimated. \citet{munoz06a} compared the colour-magnitude diagrams of Boo and UMa, concluding that Boo must be at least twice as bright as UMa. This clearly shows that there is also a considerable uncertainty in the determination of the absolute magnitudes of the observed very faint companion satellite galaxies.

For the first few orbits, the satellites disperse only slowly, getting fainter, until they approach the state of near disruption. Then the observed mass-to-light ratio $(M/L)_{\rm obs}$ increases very rapidly. Interestingly, the models qualitatively reproduce the trend interpreted to be a signature of an universal $10^7 \unit{M_{\sun}}$ halo for all dSph satellites, indicated by the dashed lines in Figure~\ref{fig_m2l}. However, about half of the snapshot values lie significantly above the trend-lines. This can be understood since our automated, virtual observer is kind of naive. A thorough observer will carefully select the stars belonging to a satellite galaxy which will reduce unrealistic values $(M/L)_{\rm obs}$ to a more realistic regime \citep[see e.g.][]{munoz06}, but will still end up at too high $(M/L)_{\rm obs}$ ratios compared to the real mass-to-light ratio.

In Figure~\ref{fig_m2l} one can also identify a group of snapshot data-points at $\MVrbin \approx -10\unit{mag}$ that have much too large $(M/L)_{\rm obs}$ ratios, marked by lighter triangle symbols. These snapshots belong to the simulations RS1-32 and RS1-64 (while the darker symbols belong to the other six simulations), both having very radial orbits. This brings the satellites very close to the galactic disc, which is not considered in the simulations presented here. Having the additional highly non-spherical potential and the effects of disc shocking will considerably affect the properties of the satellites, also depending on the impact angle with the disc. This needs to be studied in much more detail in the future.\\

The work presented here details the evolution of gas and dark-matter free satellite galaxies. But it has to be clarified in more detail how these objects can form and how many of them can condense in tidal arms, which is beyond the scope of this paper. 
\citet{wetzs05} identified at least five very massive objects in their SPH simulation, the largest one, at the tip of the tidal tail, has a mass corresponding to $\approx 3.5 \times 10^8 \unit{M_{\sun}}$ if the progenitor galaxy is scaled to the size of the MW. \citet{bourn06} found a mean production rate of somewhat more than six massive substructures ($>10^8 \unit{M_{\sun}}$) formed in the tidal tails in their simulations. 20 per~cent of these objects survive for more than 2~Gyrs. Of the non-surviving ones about two-third fall back onto their host galaxies. The other one-third was considered in their simulation as destroyed because they dropped below a mass threshold of $10^8 \unit{M_{\sun}}$. We argue that these objects should not be considered a priori as being completely destroyed. Instead, we have shown here that satellite galaxies of much lower mass may be identified on the sky and appear similar to today's known dSph galaxies of the Milky Way. This leaves room for a number of tidal dwarf remnants, even for a low number of major interaction events. However, as already noted by the authors, \citeauthor{bourn06} did not account for the larger gas fraction of galaxies in the early universe, while \citeauthor{wetzs05} used a gas fraction of 30 per~cent, despite the different treatment of the gas (sticky particles vs.\ SPH) a further major difference in the models. A higher gas fraction may enhance the efficiency of TDG production. From the current simulations it is also unclear how many lower mass ($< 10^8\unit{M_{\sun}}$) TDG candidates form, but observations show that even today's interactions can form of the order of a dozen TDGs from a single encounter (for example in AM~1353~272, \citealp{weilb00}).

If some of the dSph galaxies of the MW were of tidal origin their genesis must have happened at an early time. In all well-studied dwarf spheroidals an old stellar population is present \citep[e.g.][]{grebe04,tolst04}, they do follow a metallicity-luminosity relation (e.g.\ \citealp{mateo98,vdber99}; \citealp*{grebe03}) and have extended star formation histories \citep[e.g.][]{grebe00,ikuta02}. The latter two findings require that the galaxies maintained a considerable amount of gas for some time to continue star formation or later%
turn it on again, %
and to retain the produced metals. Today they are almost gas free \citep{galla94} and do not show ongoing star formation. As such they must have developed from gas-rich galaxies to today's dSph (\citealp{mayer01b,paset03}, but see also \citealp{grebe03}). For a given metallicity, dSph galaxies have a luminosity 10 to 100 times lower than dIrr galaxies \citep{grebe03}, being consistent with initial TDGs that form from pre-enriched material losing 90--99 per~cent of their mass as a result of tidal forces \citep{kroup97}.

A common ancient stellar population in satellite galaxies of tidal origin can be explained by considering that tidal dwarfs do not solely consist of stars born in star formation events during and after their genesis, but additionally also of old stars from the progenitor galaxy. \citet{elmeg93} argued that a newly born TDG consists of about 40 per cent old stars. Similarly, \citet{wetzs05} found that about 30 per cent of the mass of their most massive TDG candidate is in old stars, while being basically void of dark-matter. Very recently \citet{helmi06} compared the metal-poor tails of the metallicity distribution of the MW halo and of four MW dSph satellites. While the distributions were compatible among the dSph galaxies, they significantly differ compared to the MW halo. \citet{helmi06} conclude that the MW and the dSphs must have had different progenitors, since the dSphs are lacking very metal poor stars. If dSphs were of tidal origin, their oldest stars would consist of stars already pre-enriched in the progenitor galaxy.

It is also noteworthy that despite the fact that all dSph galaxies are subject to very similar boundary conditions in the Local Group they show a great diversity in their star formation histories \citep{grebe00}. If they were primordial dark-matter halos filled with gas there must have been some sporadic external or internal events which triggered star-formation from time to time in the individual galaxies.
It is also remarkable that a number of the dSph galaxies show internal structure or off-centred and twisted isophotes \citep[e.g.][]{palma03,walch03,cioni05} which is not expected if they are shielded by a massive dark-matter halo. 
A further challenge of the notion of the suggested universal minimum mass of some $10^7\unit{M_{\sun}}$ \citep{mateo98,gilmo06} would be the significant mass loss from the dark-matter halo due to the MW tides. Given the distorted inner morphology of most of the dSph satellites, why do they still have the same DM halo? Tidal mass-loss has been shown to be significant for such satellites \citep{read06}. However, this may not be a robust argument because the minimum halo-mass concerns only the mass within the observed optical light: nothing outside the kinematic data is used; so original total masses are not relevant (Gilmore priv.\ comm.). 
On the other hand, a TDG population may very easily show very different evolution histories and internal structures, even if they were born in the same environment (Figures~\ref{fig_triple}--\ref{fig_m2l}) as they are not shielded by a massive DM halo.

\section{Concluding remarks}\label{sec_conclusions}
The results presented here are a consequence of Newtonian dynamics under the assumption that the Milky Way has a massive dark-matter halo. Assuming the MW and M31 were assembled from smaller gaseous proto-galaxies, TDGs will likely have been born in this early build-up epoch.

Here we place typical TDGs into a MW-type host dark-matter halo and find that after a Hubble-time the TDGs appear very similar to dSph satellites for a large variety of orbits. However the understanding of the end stages needs to be improved by higher-resolution simulations thereby addressing the scatter of the model satellites evident in Figs.~\ref{fig_triple}--\ref{fig_m2l}. Also, the influence of the gravitational potential and shocking of the galactic disc needs to be studied.
A complete picture of TDG formation and evolution may only be achieved if high-resolution simulations of DM, stars, and gas are combined with suitable star-formation and stellar feedback models.

\emph{If} some of the Milky Ways satellite galaxies are of tidal origin, very careful inspection may be needed to identify them: they would contain old stars \citep{elmeg93,wetzs05}, may have had long lasting star-formation epochs \citep{hunte00} and appear very similar as today's dSph galaxies (\citealp{kroup97}; this work). One way to identify dSphs of a common tidal origin is to look for strong correlations of satellite orbits \citep{lynde76,palma02,metz06}, an entirely independent line-of-evidence.

As a final cautionary note: it may be useful to keep in mind that \emph{if} Newtonian dynamics is not the correct physical description for galactic problems \citep[e.g.][]{mateo98,famae05} then the overall picture considered here would remain correct (dSph may be old TDGs), but the interpretation of the dSph galaxies as objects in dynamical equilibrium \citep[e.g.][]{brada00} or as phase-space remnants (this work) would need revision.

\vspace{5mm}
\noindent{\bf Acknowledgements}\\
We thank Gerry Gilmore for helpful suggestions. PK thanks the IoA for supporting his summer-residence in Cambridge in 2005 and 2006, where this work was initiated. We are grateful to Dan Zucker for sending us the observational content of Fig.~\ref{fig_hlr}, and we thank Wyn Evans, Mark Wilkinson, Vasily Belokurov, Dan Zucker and Michael Fellhauer for very stimulating discussions at the IoA.

\bibliographystyle{mn2e}
\bibliography{}

\label{lastpage}

\end{document}